\documentclass[a4paper, amsfonts, amssymb, amsmath, reprint, showkeys, nofootinbib, twoside]{revtex4-1}
\usepackage[english]{babel}
\usepackage[utf8]{inputenc}
\usepackage[colorinlistoftodos, color=green!40, prependcaption]{todonotes}
\usepackage{amsthm}
\usepackage{xcolor}
\usepackage{graphicx}
\usepackage[T1]{fontenc}
\usepackage[pdftex, pdftitle={Article}, pdfauthor={Author}]{hyperref} % For hyperlinks in the PDF
\usepackage{scalerel}
\usepackage{tikz}
\usetikzlibrary{svg.path}

\definecolor{orcidlogocol}{HTML}{A6CE39}
\tikzset{
  orcidlogo/.pic={
    \fill[orcidlogocol] svg{M256,128c0,70.7-57.3,128-128,128C57.3,256,0,198.7,0,128C0,57.3,57.3,0,128,0C198.7,0,256,57.3,256,128z};
    \fill[white] svg{M86.3,186.2H70.9V79.1h15.4v48.4V186.2z}
                 svg{M108.9,79.1h41.6c39.6,0,57,28.3,57,53.6c0,27.5-21.5,53.6-56.8,53.6h-41.8V79.1z M124.3,172.4h24.5c34.9,0,42.9-26.5,42.9-39.7c0-21.5-13.7-39.7-43.7-39.7h-23.7V172.4z}
                 svg{M88.7,56.8c0,5.5-4.5,10.1-10.1,10.1c-5.6,0-10.1-4.6-10.1-10.1c0-5.6,4.5-10.1,10.1-10.1C84.2,46.7,88.7,51.3,88.7,56.8z};
  }
}

\newcommand\orcidicon[1]{\href{https://orcid.org/#1}{\mbox{\scalerel*{
\begin{tikzpicture}[yscale=-1,transform shape]
\pic{orcidlogo};
\end{tikzpicture}
}{|}}}}
\begin{document}
\title{Exploring thermodynamics inconsistencies in unimodular gravity: a comparative study of two energy diffusion functions}

\author{Miguel Cruz$^1$\orcidicon{0000-0003-3826-1321}}
\email{miguelcruz02@uv.mx}

\author{Norman Cruz$^{2,3}$\orcidicon{0000-0002-0737-3497}}
\email{norman.cruz@usach.cl}

\author{Samuel Lepe$^4$\orcidicon{0000-0002-3464-8337}}
\email{samuel.lepe@pucv.cl}

\affiliation{$^1$Facultad de F\'{\i}sica, Universidad Veracruzana 91097, Xalapa, Veracruz, M\'exico, \\
$^2$Departamento de F\'isica, Universidad de Santiago de Chile, Avenida Ecuador 3493, Santiago, Chile,\\
$^3$Center for Interdisciplinary Research in Astrophysics and Space Exploration (CIRAS), Universidad de Santiago de Chile, Av. Libertador Bernardo O'Higgins 3363, Estación Central, Chile,\\
$^4$Instituto de F\'{\i}sica, Facultad de Ciencias, Pontificia Universidad Cat\'olica de Valpara\'\i so, Avenida Brasil 2950, Valpara\'iso, Chile 
}

\date{\today} % Leave empty to omit a date

\begin{abstract}
In this work we study the thermodynamics formulation for unimodular gravity under the election of two different models for the energy diffusion function. Such function encodes the current for the non-conservation of the energy-momentum tensor and is usually termed as $Q(t)$. In analogy to the cosmological scenario where the cosmic expansion is influenced by $Q(t)$, the thermodynamics implications in this scheme are also determined by the choice of the function $Q(t)$, as we discuss in the work. Specifically, we consider the barotropic and the continuous spontaneous localization models as energy diffusion functions, commonly used in the literature as viable candidates to face the well-known $H_{0}$ tension. The consistency conditions demanded for the entropy of the system in terms of the cosmological parameters of the model: positive production ($dS/dt>0$) and convexity condition ($d^{2}S/dt^{2} <0$), are investigated. We show that these conditions strongly constraint the viability of both models. Additionally, we comment about our results and compare with those obtained in recent works where the restriction of the parameters for these two diffusion models was implemented with the use of cosmological data.  
\end{abstract}


\maketitle

\section{Introduction}

Despite its successful in explain many of the cosmological data over the last decades, the $\Lambda$CDM model faces increasing tensions from an observational perspective, which have exacerbated in recent years. Among the most remarkable ones we found the Hubble tension, where exist a discrepancy of the Hubble constant by measuring their value from the receding velocity of SNe Ia in a model-independent and local estimation~\cite{Riess:2016jrr,Riess:2018byc,Riess:2019cxk} and extrapolating the data coming from the CMB, assuming the $\Lambda$CDM model~\cite{Aghanim:2018eyx}. This tension is nowadays at $5\sigma$ of confidence level (CL) ~\cite{Riess:2021jrx,Addison:2017fdm,Macaulay:2018fxi,Wong:2019kwg}. On the other hand, recent observations from the James Webb Space Telescope (JWSP) have shown massive galaxy sources at $z > 7-10$, which within the standard model do not have enough time to form~\cite{liu2022accelerating}.\\

The above tensions has been addressed in different contexts that include modifications in the dark sector or in alternatives theories to general relativity (GR). One of these  alternatives to GR is Unimodular gravity (UG), originally proposed by Einstein himself~\cite{Einstein:1919gv}, and which we will explain in more details in the next section. Since it avoids introducing new fields or higher dimensions is considered an appealing approaches to address the cosmological constant (CC) problem. A remarkable consequence of the restricted diffeomorphism invariance of UG is the phenomenon of energy diffusion, which can be characterized by an specific energy diffusion function (EDF)\footnote{See for instance Ref. \cite{merced}, where a fully diffeomorphism-invariant action principle is presented for UG in the context of $BF$-type theories}. One of this EDF is known as continuous spontaneous localization (CSL) model~\cite{Pearle:1976ka,Ghirardi:1985mt,Pearle:1988uh,Ghirardi:1989cn}, in which energy is created when quantum collapse occurs~\cite{Pearle:1994rj}. This process is assumed to be small but cumulative along the cosmic history and leads to the appearance of an effective CC that drives the cosmic acceleration and in which the contribution arising from violations of the energy-momentum conservation is encoded by means of the EDF~\cite{Josset:2016vrq}. Moreover, it modifies the scaling behaviour of matter and it provides a sensitive prediction for current precision measurements. Noteworthy, as shown in Ref.~\cite{Corral:2020lxt}, diffusion cancels out the appearance of higher-order equations that might arise in cosmological contexts~\cite{garcia2019cosmic}, alleviating an eventual ill-posedness of their associated Cauchy initial-value problem. The observational implications of this setup has been studied in~\cite{Corral:2020lxt} by means of SNe Ia and Hubble data, showing that CSL model is incompatible at the background level with previous laboratory experiments~\cite{torovs2017colored}.\\

Other EDF in cosmological UG models has been explored using a $Q$ function described by the Barotropic Model (BM), where $Q := \alpha \rho$ ($\alpha$ is a constant and $\rho$ represents the energy density of the main fluid)~\cite{Corral:2020lxt} and a $Q$ which  considers third order derivatives in the scale factor, related to a Jerk parameter, used to describe the late accelerated expansion without the inclusion of a DE term~\cite{garcia2019cosmic}.
Interaction between DM and DE in the framework of UG to address the the $H_0$ tension has been investigated in ~\cite{Perez:2020cwa,LinaresCedeno:2020uxx}. 

To face the challenge from JWST's observations has been to consider a string-inspired scenario where the DE sector consists of a negative CC and a evolving component with positive energy density on top, whose EoS is allowed to cross the phantom divide, changing drastically the growth of structure compared to the $\Lambda$CDM model ~\cite{Adil_2023}.\\ 

Interestingly, in the CSL model explored in~\cite{Corral:2020lxt}, in the very far future, the model displays the feature of a variable effective CC which appears with sign flip opening the possibility of having a phase transition from de Sitter to anti de Sitter expansion. We obtain then a source of accelerated expansion that decays with the cosmic evolution, as in the above proposal. Indeed, the absence of turning points in the Hubble parameter, as predicted by $\Lambda$CDM and supported by experiments, is compatible with models with EDF possessing a sign flip in the variable effective CC~\cite{Corral:2020lxt} in the very far future. This may be an attractive consequence regarding the swampland criteria on cosmology~\cite{Agrawal:2018own}. Although different scenarios with variable CC have been proposed beyond $\Lambda$CDM~\cite{Caldwell:1997ii,Clifton:2011jh,vanPutten:2015wma,Colgain:2018wgk,10.1093/mnrasl/slz158,NOJIRI20171}, diffusion provides a rich phenomenology dispensing from additional fields and higher-curvature corrections to the Einstein--Hilbert action. 

Despite the above promissory results found in cosmological UG models, their thermodynamics consistency is an open question to explore, since the EDF implies that the entropy is not constant, on the contrary to the $\Lambda$ model. In other scenarios that predicts also energy diffusion exhibit some drawbacks under a thermodynamic exploration. For instance, in Rastall's theory~\cite{rastall1976theory} the non-conservation of the energy-momentum tensor is given as $\nabla^{\mu}T_{\mu \nu} = \zeta \nabla^{\mu}(g_{\mu \nu}R)$ with $\zeta$ being the Rastall parameter, such condition is contained in the UG framework, as we will see later. Within the Rastall scheme the thermodynamic perspective provides a strong criterion to construct the EDF in order to obtain a well defined model from the thermodynamics point of view~\cite{cruz2019thermodynamics}.

In the case of the renormalizable theory of Ho\v{r}ava-Lifshitz gravity~\cite{hovrava2009quantum}, a EDF was explored as a mechanism for transferring energy between the bulk and the boundary of spacetime, showing that through the cosmic evolution the second law can be verified only some narrow region of the redshift. Besides a clear indication of non-equilibrium between bulk and boundary was found implying that the thermal equilibrium can be explored further \cite{cruz2015bulk}.\\

For a dissipative dark sector exists also a non conservation of the energy densities of these kind of fluids. In the framework of the causal\cite{I.S.1,I.S.2} and non-causal~\cite{Eckart} approaches these non-perfect fluids lead to  interesting behavior like a phantom DE, unified models without DE, possible new future singularities or non big bang singularity~\cite{Cataldo:2005qh,Cruz:2017bcv,Cruz:2017lbu,Cruz:2016rqi,PhysRevD.100.083524, Cruz_2018,cruz2023non,C_rdenas_2020}, and also a way to face the $H_{0}$ tension~\cite{tensionH0,BulktensionH} or the $\sigma_{8}-\Omega_{m}$ tension~\cite{remedyforplankanlssdata}. Nevertheless the evaluation of the thermodynamics conditions, like, for example, the entropy production associated to each of the above scenarios or near equilibrium condition when dissipation is present, imposes thigh constraints on the range of the parameters which characterized them, or even discard some. Even more, a further research is related to a recent and very interesting results that were found in ~\cite{Cruz:2023xjp}, showing that there are $P-V$ transitions in Einstein gravity considering noninteracting fluids, cold dark matter and holographic DE. The novelty is that Einstein gravity does not support $P-V$ phase transitions if only one fluid is present, but may occur in modified gravity  ~\cite{Kong_2022}.\\ 

The aim of the this paper is to explore the thermodynamics behavior of a cosmological UG model with the barotropic and CLS as EDF, regarding the fulfillment of the following three main conditions: i) positivity of the entropy, $S>0$ ; ii) second law $dS/dt\geqslant 0$, iii) convexity (tendency to balance) $d^{2}S/dt^{2}<0$.  All these aspects must be explored in detail to ensure the physical viability of these EDF, and the constraints that the above conditions impose on the associated.\\ 

This paper is organized as follows: Section \ref{sec:diff} contains a concise description of UG, the background dynamics equations of motion are established in the context of a Friedmann-Lema\^itre-Robertson-Walker (FLRW) spacetime. In Section \ref{sec:thermo} the thermodynamics formulation of UG is discussed with the consideration of two different diffusion functions well-known in the literature. Based on the values of the cosmological parameters of the model, we explore the inconsistencies that emerge in both scenarios. We also comment about these physical inconsistencies and some recent results where cosmological data have been used for this model. Section \ref{sec:final} is devoted to the final comments of our work. Units are chosen so that $8\pi G=c=k_{B}=1$ throughout this manuscript.  %\cite{corral2020diffusion} 

%%%%%%%%%%%%%%%%%%%%%%%%%%%%%%%%%%
\section{Cosmological model with diffusion}
\label{sec:diff}
%%%%%%%%%%%%%%%%%%%%%%%%%%%%%%%%%

In this section we provide a brief description of UG and we establish the dynamical equations at background level for this model. The cosmological scenario will be described assuming a FLRW metric with flat space section ($k=0$).\\ 

As in GR, it is well-known that UG can be described using the Einstein-Hilbert Lagrangian but through the introduction of a constraint which fixes to a constant the determinant of the metric \cite{10.1119/1.1986321}, such consideration lead to the following set of equations of motion for the gravitational field after solving for such Lagrange multiplier\footnote{The variation of the unimodular action with respect to metric leads to the following equations of motion
\begin{equation*}
    R_{\mu \nu} - \frac{1}{2}g_{\mu \nu}R + \lambda(x) g_{\mu \nu}  = T_{\mu \nu}, %\label{eq:unieom}
\end{equation*}
and the variation with respect to the Lagrange multiplier, termed as $\lambda(x)$, leads to $\sqrt{-g} = f$, which is the well-known unimodular condition, being $f$ a constant.}
\begin{equation}
     R_{\mu \nu} - \frac{1}{4}g_{\mu \nu}R  = T_{\mu \nu}- \frac{1}{4}g_{\mu \nu}T, \label{eq:unieom}
\end{equation}
these dynamical equations represent the trace-free part of GR equations and $T_{\mu \nu}$ is the energy-momentum tensor of the matter content, as usual. Now, by applying the Bianchi identity to Eq. (\ref{eq:unieom}), one gets 
\begin{equation}
    \nabla^{\mu}T_{\mu \nu} = \frac{1}{4}\nabla^{\mu}[g_{\mu \nu}(R+T)]=\frac{1}{4}\partial_{\nu}(R+T), \label{eq:compare}
\end{equation}
therefore the energy momentum-tensor is no longer locally conserved. As a manifest relation between UG and Rastall theory, for an ultra-relativistic gas satisfying the condition, $T = 0$, the non-conservation condition considered by Rastall~\cite{rastall1976theory} is obtained from the equation given above by setting the value $1/4$ for the Rastall parameter.\\ 

Some comments are in order. Since the equations of motion (\ref{eq:unieom}) can be written alternatively as
\begin{equation}
    R_{\mu \nu} - \frac{1}{2}g_{\mu \nu}R + \frac{1}{4}(R+T)g_{\mu \nu} = T_{\mu \nu}, \label{eq:unieom2}
\end{equation}
thus Lagrange multiplier introduced in the UG action is identified as $\lambda = (1/4)(R+T)$. Notice that if the condition of conservation for the energy-momentum tensor is imposed, we can obtain from equation (\ref{eq:compare}) the following result, $R+T = \mbox{const.}$; if we identify such constant as $4\Lambda$ thus Eq. (\ref{eq:unieom2}) represents the GR equations of motion with CC. On the other hand, the Eq. (\ref{eq:compare}) allows us to provide an explicit expression for the Lagrange multiplier, yielding
\begin{equation}
    \lambda(x) = \Lambda + \int_{l} J(x), \label{eq:multiplier}
\end{equation}
where $\Lambda$ is a constant of integration which in principle can take positive or negative values, therefore the origin of this constant does not come from a Lagrangian with cosmological term, this is a formal difference between GR and UG, besides we have defined the current $J_{\nu} := \nabla^{\mu}T_{\mu \nu}$ for the non-conservation of the energy-momentum tensor which must be integrated in an arbitrary path $l$. The use of the above results allow us to write Eq. (\ref{eq:unieom2}) alternatively as 
\begin{equation}
    R_{\mu \nu} - \frac{1}{2}g_{\mu \nu}R + \left( \Lambda + \int J(x) \right) g_{\mu \nu}  = T_{\mu \nu}. \label{eq:unieom3}
\end{equation}
From these equations of motion is possible to identify the Lagrange multiplier as an {\it effective cosmological constant.} With the advantage that the emerging constant of integration can play the role of the CC allows to UG to solve the fundamental issue of vacuum energy. Due to the homogeneity and isotropy of the universe at cosmological scales, the expression (\ref{eq:multiplier}) for the effective CC must be a function of the cosmic time only
\begin{equation}
    \lambda_{\mathrm{eff}}(t) = \Lambda + \int_{l} J(t),
\end{equation}
usually the second term appearing in the last expression is called as the EDF and generally denoted as, $Q(t) := \int_{l} J(t)$, then the form of the effective CC is simply
\begin{equation}
    \lambda_{\mathrm{eff}}(t) = \Lambda + Q(t).
     \label{eq:time}
\end{equation}
Therefore this formalism depends on explicit proposals for  $Q(t)$ in order to solve the background dynamics of the model. According to the sign of the integration constant $\Lambda$ and the behavior of the EDF, $Q(t)$, sign flip in $\lambda_{\mathrm{eff}}(t)$ is allowed, as commented previously.\\ 

In the context of a FLRW spacetime, the Friedmann equations for UG obtained from Eq. (\ref{eq:unieom3}) take the following form assuming that matter content is described by a perfect fluid characterized by its energy density $\rho$, and pressure $p$, yielding
\begin{align}
& 3H^2 = \rho + \lambda_{\mathrm{eff}}(t) = \rho + \Lambda + Q, \label{eq:fried1}\\
& 2\dot{H} + 3H^2 = -p+\lambda_{\mathrm{eff}}(t)= -(p - Q) + \Lambda, \label{eq:accel}
\end{align}
together with
\begin{equation}
\dot{\rho} + 3H(\rho + p)= - \dot{Q},\label{eq:nonconsrho}
\end{equation}
where we have considered the condition given in (\ref{eq:time}). The presence of the EDF, $Q(t)$, in the set of equations given above is crucial to distinguish UG from GR plus CC, the sign of the integration constant $\Lambda$ is arbitrary and the continuity equation (\ref{eq:nonconsrho}) now has a source term which allows to make contact with other cosmological models as we comment below. The existence of $Q(t)$ in the set of equations (\ref{eq:fried1})-(\ref{eq:nonconsrho}) resembles the background dynamics associated to more general cosmological scenarios as bulk viscosity and matter creation; such schemes allow dissipative processes which induce deviations from the equilibrium pressure of the fluid due to the existence of viscous pressure and interchange of energy between spacetime and the matter sector activating the particle production. Both effects can be characterized by means of a pressure term distinct from the pressure of the fluid, therefore the diffusion introduced by means of the function $Q(t)$ in UG contributes to the cosmic expansion as bulk viscosity and the rate of particle production does in the aforementioned cosmological scenarios, for more details see for instance~\cite{Cataldo:2005qh,Cruz:2017bcv,Cruz:2017lbu,Cruz:2016rqi,PhysRevD.100.083524, Cruz_2018,cruz2023non,C_rdenas_2020,tensionH0,BulktensionH,remedyforplankanlssdata,us1_2024,us2_2024}. As usual, the dot denotes derivatives with respect to cosmic time and $H$ is the Hubble parameter defined in terms of the scale factor $a(t)$.

%%%%%%%%%%%%%%%%%%%%%%%%%%%%%%%%%%%%%%%%%%%%%%%%
\section{Thermodynamics discussion}
\label{sec:thermo}
%%%%%%%%%%%%%%%%%%%%%%%%%%%%%%%%%%%%%%%%%%%%%%%%
In order to discuss some implications of the UG's thermodynamics formulation, we will focus on the BM, in which the EDF is described by
\begin{equation}
    Q:=\alpha \rho, \label{eq:alfa} 
\end{equation}
which represents the most simple assumption about this term, as can be seen in Refs. \cite{Corral:2020lxt, LinaresCedeno:2020uxx}. If we take into account the change of variable $t\rightarrow z$ being $z$ the cosmological redshift related to the scale factor by means of the standard relation $a(t)=(1+z)^{-1}$, then for an arbitrary function $Q$ the integration of Eq. (\ref{eq:nonconsrho}) in terms of the redshift and $p=0$, leads to
\begin{equation}
    \rho(z)=\rho_{0}(1+z)^{3}\left[1-\frac{1}{\rho_{0}}\int \frac{dQ}{(1+z)^{3}}\right],
\end{equation}
indicating that $\mbox{dim}(\rho) = \mbox{dim}(Q)$, thus the definition of the EDF (\ref{eq:alfa}) is consistent. From now on the subscript zero denotes evaluation of cosmological quantities at present time, $z=0$.\\

On the other hand, inserting the EDF (\ref{eq:alfa}) in the Eq. (\ref{eq:nonconsrho}), yields
\begin{equation}
    \rho(z)= \rho_{0}(1+z)^{3(1+\omega_{\mathrm{eff}})},\label{eq:energydensity}
\end{equation}  
where we also considered a barotropic equation of state between the energy density of the fluid and its pressure, $p=\omega \rho$, with constant parameter state $\omega$ and we have defined the following effective parameter state
\begin{equation}
    \omega_{{\mathrm{eff}}} := \frac{\omega-\alpha}{1+\alpha}.\label{eq:17}
\end{equation}
Notice that the solution for the energy density given in Eq. (\ref{eq:energydensity}) is simply obtained from the redefinition, $\rho := (1+\alpha)\rho$. In this case the positivity of the energy density is guaranteed for $\alpha > -1$. However, as we will discuss later, the interval $\alpha > 0$ can lead to thermodynamics inconsistencies, as negative entropy. For $\alpha = 0$ the energy density (\ref{eq:energydensity}) recovers the usual behavior of a dark energy component with constant parameter state $\omega$ known as $\omega$CDM model \cite{wcdm}. On the other hand, the case $\omega = \alpha$ leads at effective level to a behavior that is consistent with the dark matter description since obeys a pressureless equation of state, $\omega_{\mathrm{eff}}=0$. According to the discussion given previously for the UG, the normalized Hubble parameter takes the form
\begin{equation}
    E(z):=\frac{H(z)}{H_{0}} = \sqrt{\Omega_{\Lambda}+\Omega_{0}(1+z)^{3(1+\omega_{\mathrm{eff}})}}, \label{eq:normalized}
\end{equation}
being $\Omega$ the fractional energy density associated to the component $\rho$ and $H_{0}$ is the Hubble constant. Notice that the normalization condition obtained from the previous expression takes the explicit form
\begin{equation}
    \Omega_{\Lambda}+(1+\alpha)\Omega_{0}=1, \label{eq:normalization}
\end{equation}
which leads to the condition 
\begin{equation}
    \alpha = \frac{1-\Omega_{\Lambda}}{\Omega_{0}}-1,
\end{equation}
therefore the $\alpha$-parameter is restricted to the following half-open interval $(-1, (1-\Omega_{\Lambda})\Omega_{0}^{-1}-1]$, in order to fulfill the positivity condition of the energy density (\ref{eq:energydensity}) and to guarantee the normalization condition (\ref{eq:normalization}) with $0<\Omega_{\Lambda}, \Omega_{0} < 1$. For $\alpha=0$ we recover the standard case $ \Omega_{\Lambda}+\Omega_{0}=1$ for the normalization condition (\ref{eq:normalization}). Also from the normalized Hubble parameter, the deceleration parameter can be constructed by transforming its definition in terms of time, $q(t)=-1-\dot{H}/H^{2}$, to a redshift function as follows
\begin{equation}
    q(z)=-1+(1+z)\frac{E'(z)}{E(z)}=-1+\frac{3}{2}\frac{1+\omega}{\left(1+\alpha +\frac{\Omega_{\Lambda}}{\Omega(z)}\right)}, \label{eq:deceleration}
\end{equation}
then at present time this model exhibits a quintessence behavior $q_{0} > -1$ and at the far future a de Sitter expansion is recovered, $q(z=-1)=-1$. This represents an interesting behavior of UG in the light of DESI results where the quintessence scenario is strongly favored for dynamical dark energy models \cite{desi}. From Eq. (\ref{eq:17}) we can observe that at effective level the crossing to the phantom regime ($\omega_{\mathrm{eff}} < -1$) is not allowed even for the case $\omega=0$. Thus the $\alpha$-parameter determines the type of cosmic evolution in this cosmological model. More details about the cosmological implications of this model can be found in appendix \ref{sec:app}.\\ 

The Gibbs equation is given as \cite{callen}
\begin{equation}
    TdS = dU+pdV, \label{eq:gibss}
\end{equation}
being $U$ the internal energy defined as, $U:=\rho V$, and $V\propto a^{3}(t) = V_{0}(1+z)^{-3}$ is the volume. Then, using Eq. (\ref{eq:nonconsrho}) in terms of the redshift together with Eq. (\ref{eq:gibss}) one gets the following expression for the differential of the entropy 
\begin{equation}
    TdS =-VdQ, \label{eq:entropy}
\end{equation}
as can be seen, the EDF plays a relevant role in the thermodynamics context since catalyzes the entropy production. Thus the cosmological expansion in this model can be seen as an irreversible process, which provides a more realistic physical model since deviations from equilibrium are expected. According to Eq. (\ref{eq:alfa}) the previous expression is simply given as
\begin{equation}
    TdS =-\alpha V_{0}(1+z)^{-3}d\rho. \label{eq:entropy2}
\end{equation}
On the other hand, the implementation of the effective temperature method explored in \cite{grandon} allows us to compute the temperature of the fluid. Assuming that the fluid satisfies number conservation together with the Eq. (\ref{eq:17}), then we can write the evolution equation of the temperature as follows \cite{maartens}
\begin{equation}
    \dot{T} = -3HT\omega_{\mathrm{eff}},
\end{equation}
where the equation of state of the fluid is now mediated by the effective parameter state. In terms of the redshift the temperature obtained from the above equation takes the form
\begin{equation}
    T(z)=T_{0}(1+z)^{3\omega_{\mathrm{eff}}},\label{eq:tempe}
\end{equation}
this is the typical behavior of the positive definition of temperature obtained in the single fluid description \cite{maartens}. As commented above the case $\omega=\alpha$ represents the dark matter description at effective level since $\omega_{\mathrm{eff}}=0$, then in this case the temperature of the fluid remains constant, in agreement with the thermodynamics description results of standard cosmology. The $\omega \neq 0$ model for the dark matter description was proposed as a generalized model, see Ref. \cite{hu}; according to some extensive analysis of this generalized dark matter model the condition, $\omega > 0$, was found to be consistent through the cosmic history and supported by cosmological data \cite{extensive, extensive1}. The derivative of entropy in terms of the redshift can be computed by inserting the Eqs. (\ref{eq:energydensity}) and (\ref{eq:tempe}) into the expression (\ref{eq:entropy2}), then
\begin{equation}
    \frac{dS(z)}{dz} = -\frac{3\alpha U_{0}}{T_{0}}(1+\omega_{\mathrm{eff}})(1+z)^{-1}, \label{eq:derivative}
\end{equation}
where $U_{0}$ is the internal energy at present time given as, $U_{0}:=\rho_{0}V_{0}$. As case of cosmological interest, we focus at first place on the value $\omega=0$, therefore we simply have 
\begin{align}
    & \rho(z)= \rho_{0}(1+z)^{\frac{3}{1+\alpha}},\\
    & U(z) = U_{0}(1+z)^{-\frac{3\alpha}{1+\alpha}},\\
    & \frac{dS(z)}{dz} = -\frac{3U_{0}}{T_{0}}\left(\frac{\alpha}{1+\alpha} \right)(1+z)^{-1}, \label{eq:first}
\end{align}
the second derivative of entropy is computed straightforwardly from the above result, yielding
\begin{equation}
    \frac{d^{2}S(z)}{dz^{2}} = \frac{3U_{0}}{T_{0}}\left(\frac{\alpha}{1+\alpha} \right)(1+z)^{-2}. \label{eq:second}
\end{equation}
The evolution of natural processes in physical systems, i.e., their tendency to reach thermodynamical equilibrium, demand two simultaneous consistency conditions on the derivatives of the entropy, we must have positive entropy production $dS/dt > 0$ and the convexity condition given by $d^{2}S/dt^{2} < 0$, which in turn result as $dS/dz < 0$ and $d^{2}S/dz^{2} > 0$ when the change of variable $t\rightarrow z$ is considered.\\ 

Regarding the conditions demanded for the entropy we discuss two different cases dictated by the values of the parameter $\alpha$: $-1 < \alpha < 0$ and $\alpha > 0$. It is worthy to note that according to Eqs. (\ref{eq:first}) and (\ref{eq:second}), both consistency conditions can not be satisfied simultaneously in the interval $-1 < \alpha < 0$. On the other hand, for $\alpha > 0$ such conditions are fulfilled, however, this interval excludes the region $-1 < \alpha < 0$ in which the positivity of the energy density (\ref{eq:energydensity}) is also guaranteed. This situation represents a severe problem in the thermodynamics formulation of UG under the barotropic assumption for $Q$ given in (\ref{eq:alfa}), since the second law can be violated in some region of its parameters space. The fulfillment of the second law is guaranteed at large scales as reported in \cite{pavon1}, thus all cosmological models must preserve it in any case, see also Ref. \cite{pavonk}. Notice that from equation (\ref{eq:first}), the entropy is simply given as 
\begin{equation}
    S(z) = S_{0}-\frac{3U_{0}}{T_{0}}\left(\frac{\alpha}{1+\alpha} \right)\ln (1+z). \label{eq:positivity}
\end{equation}
From the above result we obtain that the Nernst theorem (positivity of entropy) \cite{callen} is only guaranteed along the cosmic evolution if we consider the interval $-1 < \alpha < 0$. However, this interval for the $\alpha$-parameter leads to the violation of both consistency conditions demanded for the entropy as discussed previously. It is clear that for the diffusion function chosen the fulfillment of the thermodynamic second law implies the main cosmic fluid must gaining a positive flux of energy during the cosmic evolution. If we consider the case $\omega=\alpha$ the effective parameter state takes the value $\omega_{\mathrm{eff}}=0$, then Eq. (\ref{eq:derivative}) is simply 
\begin{equation}
    \frac{dS(z)}{dz} = -\frac{3\alpha U_{0}}{T_{0}}(1+z)^{-1}. \label{eq:derivative2}
\end{equation}
In consequence, and as discussed previously, both consistency conditions for the entropy are satisfied simultaneously for $\alpha>0$ only. A direct integration of this latter result leads to the same conclusions of those around Eq. (\ref{eq:positivity}): the Nernst theorem together with the second law of thermodynamics cannot be satisfied in this scenario of UG even under the consideration of a viable generalization in the description of the dark matter sector. The thermodynamics consistency in both cases discussed is recovered for $\alpha=0$, which implies $dS=0$ leading to $S(z)=S_{0}$, i.e., a barotropic cosmic expansion described as a reversible process in which the energy is conserved.\\

Another case generally found in the context of UG is the proposal for the EDF known as the CSL model, see for instance Ref. \cite{csl, Josset:2016vrq}. This model leads to violation of energy-momentum conservation and consists in non-unitary modifications of quantum dynamics. Born's rule for probabilities of experimental outcomes is recovered by involving non-linearity and stochasticity described by a Markovian evolution equation for the density matrix $\hat{\rho}$. This kind of equation has been widely used in the context of black holes physics to explain phenomena such as creation or evaporation of black holes and Hawking information puzzle \cite{bh1, bh2, bh3}, this kind of scenario explains the spontaneous collapse of wave function in quantum mechanics and has the feature of having a not constant behavior for the average energy, $\langle E \rangle = \mbox{Tr}[\hat{\rho} \hat{H}]$, being $\hat{H}$ the Hamiltonian operator. For non-relativistic particles the diffusion process of the wave function in Hilbert space is described by
\begin{equation}
    \dot{Q}=-\xi_{\mathrm{CSL}}\rho, \label{eq:csl1}
\end{equation}
being $\xi_{\mathrm{CSL}}$ the localization rate, in the cosmological context the CSL of baryons leads to an energy-momentum violation current and $\rho$ is the energy density of baryonic fluid. Using the expression given above in Eq. (\ref{eq:entropy}), one gets
\begin{equation}
    \frac{dS}{dt} = \xi_{\mathrm{CSL}}\frac{U(t)}{T(t)},\label{eq:2nd}
\end{equation}
then according to the above equation positive entropy production is guaranteed by demanding the condition $\xi_{\mathrm{CSL}} > 0$ since $U(t), T(t) >0$. The positivity condition for the parameter $\xi_{\mathrm{CSL}}$ is in agreement with the values constrained from the experiments, see Ref. \cite{torovs2017colored}, where the interval, $3.310^{-42} \ s^{-1} <  \xi_{\mathrm{CSL}} < 2.810^{-29} \ s^{-1}$, was obtained for this parameter; therefore the second law of thermodynamics can be achieved in this scenario by considering $\xi_{\mathrm{CSL}}>0$. \\

A second scenario is given in the cosmological context as \cite{csl, Josset:2016vrq}
\begin{equation}
    \dot{Q}=-\xi_{\mathrm{CS}}\rho_{0}\left(\frac{a_{0}}{a(t)}\right)^{3},  \label{eq:csl2}
\end{equation}
which describes violations of the energy-momentum tensor compatible with Lorentz invariance for massive and massless particles. In this case $\rho_{0}$ is the energy density for photons and the experimental results \cite{torovs2017colored} constrained the parameter $\xi_{\mathrm{CS}}$ within the interval $-10^{-21} \ s^{-1} <  \xi_{\mathrm{CS}} < 2\times 10^{-21} \ s^{-1}$. The negative region for $\xi_{\mathrm{CS}}$ represents an endothermic evolution in this kind of cosmological scenario. In this case we obtain from Eq.  (\ref{eq:entropy})
\begin{equation}
    \frac{dS}{dt} = \frac{a^{3}_{0}V_{0}\xi_{\mathrm{CS}}\rho_{0}}{T(t)},\label{eq:2nd2}
\end{equation}
therefore according to the result given in the latter equation, the so-called endothermic evolution ($\xi_{\mathrm{CS}}<0)$ is forbidden by thermodynamics since violates the second law. The positivity of the parameter $\xi_{\mathrm{CS}}$ must be demanded to guarantee thermodynamics consistency.\\

To end this section some comments are in order. Regarding the BM (\ref{eq:alfa}), in Ref. \cite{Corral:2020lxt} the value of the parameter $\alpha$ was inferred from a separate analysis of $H(z)$ data (OHD) and supernovae together with the joint analysis of these data sets. According with their results $\alpha$ is consistent with the condition $\alpha > 0$, but the uncertainties obtained from the OHD and supernovae data can lead to negative values. The joint data analysis results confirm that positive values for $\alpha$ are preferred. In an independent analysis, in Ref. \cite{LinaresCedeno:2020uxx} the parameter $\alpha$ was inferred using different database, specifically, CMB and late times data sets, which cover supernovae, quasars and cepheids. In this second scenario the analysis of joint data reveals again that the parameter $\alpha$ is consistent with positive values. Despite that the parameters of UG can be fitted using cosmological data, the theory is physically inconsistent since its thermodynamics formulation is plagued of inconsistencies independently of the value of the $\alpha$-parameter, besides from two physical conditions the interval $-1 < \alpha \leq 0$ is more favored than $\alpha >0$, as discussed above. The physical consistency of this model is only possible with $\alpha =0$.\\ 

On the other hand, the constrained value for $\xi_{\mathrm{CSL}}$ from the joint data analysis in Ref. \cite{LinaresCedeno:2020uxx} is negative implying an endothermic evolution, this is consistent from the quantum perspective but forbidden from the thermodynamics point of view since the second law is violated as can be seen from Eq. (\ref{eq:2nd}).

%%%%%%%%%%%%%%%%%%%%%%%%%%%%%%%%%%%%%%%%%%%%%%%%
\section{Final Remarks}
\label{sec:final}
%%%%%%%%%%%%%%%%%%%%%%%%%%%%%%%%%%%%%%%%%%%%%%%%

In this work we focused on the implications of the thermodynamics description for UG, which is characterized by the appearance of diffusive effects due to the unimodular condition; these effects induce the non-conservation of the energy-momentum tensor. In our description two different models are assumed for the diffusion term denoted as $Q(t)$, such elections are common in the literature and have been used to face, for example, the Hubble tension. For the BM some cosmological quantities are discussed, the model exhibits a quintessence behavior and depends on the values of two parameters only: $\alpha$ which determines the diffusive effects via the $Q$-term and $\omega$ that corresponds to the parameter state of the cosmic fluid relating its energy density to the pressure, as usual, both parameters are real valued for physical meaning and besides are assumed to be constants. The results emerging from the standard thermodynamics description with the aforementioned setup for UG are summarized as follows: for $\omega=0$ the second law of thermodynamics and the convexity condition for the entropy are fulfilled simultaneously only for positive values of the parameter $\alpha$ but in this case the Nernst theorem is not guaranteed. It is worthy to mention that the negative region $-1 < \alpha < 0$ is accessible due to the positivity of the energy density, however, for this interval the positive entropy production and the convexity condition are not fulfilled simultaneously. On the other hand, our formulation admits a generalized description for dark matter ($\omega \neq 0$) but as in the standard case ($\omega=0$) the thermodynamics issues of the model persist. This opens the possibility to explore some alternatives to obtain a healed thermodynamics description. For instance, the inclusion of the apparent horizon to consider a bulk-boundary interaction, see Ref. \cite{cruz2015bulk}, we leave this for future investigation.\\ 

Finally, the thermodynamics description of the CSL diffusion model given in Eq. (\ref{eq:csl1}) is well-defined for $\xi_{\mathrm{CSL}}>0$, therefore the endothermic evolution case ($\xi_{\mathrm{CSL}} < 0$) is forbidden by thermodynamics as can be seen from Eq. (\ref{eq:2nd}). On the other hand, for the model given in (\ref{eq:csl2}), the negative region of the parameter $\xi_{\mathrm{CS}}$ (endothermic evolution) is also discarded by thermodynamics. Despite the endothermic evolution is allowed in both models from the quantum perspective this is forbidden from the thermodynamics point of view.\\

As a final conclusion it can be highlighted that the thermodynamic criteria imposes severe restrictions to the EDF, which implies that further exploration is needed to construct them consistently from the thermodynamic point of view.

\section*{Acknowledgments}
MC work has been supported by S.N.I.I. (CONAHCyT-M\'exico).

\appendix
\section{Cosmological implications of the barotropic EDF}
\label{sec:app}
In this appendix we discuss briefly the cosmological implications of the model (\ref{eq:alfa}) and the interval $-1 < \alpha < 0$, which is demanded by the positivity of the energy density as mentioned before. From Eqs (\ref{eq:fried1}), (\ref{eq:accel}) and (\ref{eq:nonconsrho}) it is possible to write the general form of the deceleration parameter as follows
\begin{equation}
    q=-1+\frac{3}{2}\left(1+\frac{\omega \rho-Q-\Lambda}{\rho+Q+\Lambda} \right), \label{eq:decel}
\end{equation}
 which leads to the condition
 \begin{equation}
     \frac{\Omega_{\Lambda}}{\Omega(z)}>\frac{1}{2}-\frac{Q(z)}{\rho(z)},
 \end{equation}
in order to satisfy $-1 < q(z)<0$. For the EDF (\ref{eq:alfa}) one gets, $\Omega_{\Lambda}/\Omega(z) > 1/2 -\alpha$. As can be seen, for the interval $-1 < \alpha < 0$ the phantom regime is not accessible, as discussed previously. At the beginning of dark energy domination epoch given by the equality $\Omega_{\Lambda}=\Omega(z_{c})$ for a redshift value $z_{c}$, we obtain from the previous expression for $q(z)$
\begin{equation}
    q(z_{c})=-1+\frac{3}{2\left[2+\frac{Q(z_{c})}{\rho(z_{c})}\right]},
\end{equation}
then the decelerated-accelerated transition of the cosmic evolution is described by $Q(z_{c})=-\rho(z_{c})/2$, from where we identify $\alpha=-1/2$, which is consistent with the interval $-1 < \alpha < 0$ if we consider the model (\ref{eq:alfa}). If we now focus on the matter domination epoch ($\rho \gg \Lambda$), we can write for the deceleration parameter (\ref{eq:decel}) in this limit as follows
\begin{equation}
    \frac{Q(z)+\Lambda}{Q(z)+\rho(z)}\approx 0,
\end{equation}
which implies $\Lambda \approx -Q$. According to the barotropic model (\ref{eq:alfa}) we have in this case $\Lambda \approx -\alpha \rho(z)$, therefore $\alpha < 0$ for $\Lambda > 0$. Notice that the transition from the early matter domination epoch to late times cosmic evolution is well described by the parameter $\alpha$ lying in the interval $-1 < \alpha < 0$. 

%\begin{thebibliography}{4}
\bibliographystyle{ieeetr}
\bibliography{biblio2.bib}
%\end{thebibliography}
\end{document}